\documentclass[aip,jcp,reprint,showpacs,amsmath,amssymb,superscriptaddress]{revtex4-1}

\usepackage{graphicx}
\usepackage{bm}

\newcommand{\be}{\begin{equation}}
\newcommand{\ee}{\end{equation}}
\newcommand{\bea}{\begin{eqnarray}}
\newcommand{\eea}{\end{eqnarray}}
\newcommand{\nn}{\nonumber}

\newcommand{\rr}{\mathbf{r}}

\newcommand{\kk}{\mathbf{k}}

\newcommand{\F}{{\cal F}}
\newcommand{\rhow}{\rho(\mathbf{r})}
\newcommand{\rhoi}{\rho_i(\mathbf{r})}
\newcommand{\rhois}{\{\rho_i(\mathbf{r})\}}
\newcommand{\rhoiOs}{\{\rho_i^0\}}

\begin{document}

\title{Scalar Fundamental Measure  Theory for Hard Spheres in Three Dimensions. Application to Hydrophobic Solvation}

\author{Maximilien Levesque}
\author{Rodolphe Vuilleumier}
\author{Daniel Borgis}
\email{daniel.borgis@ens.fr}
\affiliation{\'Ecole Normale Sup\'erieure, D\'epartement de Chimie, UMR 8640 CNRS-ENS-UPMC, 24, rue Lhomond, 75005 Paris, France}

\date{\today}

\begin{abstract}
Hard-sphere mixtures provide one a solvable reference system that can be used to improve the density functional theory  of realistic molecular fluids. We show  how the  Kierlik-Rosinberg's scalar version of the fundamental measure density functional  theory of hard spheres [Phys. Rev. A, {\bf 42}, 3382 (1990)], which presents computational advantages with respect to the original Rosenfeld's vectorial formulation or its extensions, can be implemented and minimized in three dimensions to describe fluid mixtures in complex environments. This implementation is used as a basis for  defining a molecular density functional theory of water around molecular hydrophobic solutes of arbitrary shape.
\end{abstract}

\maketitle

\section{Introduction}

The numerical  methods that have emerged in the second part of the last century from  liquid-state theories\cite{hansen, gray-gubbins-vol1}, including integral equation theory in the interaction-site\cite{Chandler-RISM,hirata-rossky81,hirata-pettitt-rossky82,reddy03,pettitt07,pettitt08}
or molecular\cite{blum72a,blum72b,patey77,carnie82,fries-patey85,richardi98,richardi99} picture, classical density functional theory (DFT)\cite{evans79,evans92,Wu07}, or classical fields theory\cite{chandler93,tenwolde01,coalson96}, have become methods of choice for 
many physical chemistry or chemical engineering applications\cite{gray-gubbins-vol2,wu06}. They can yield reliable predictions for both the microscopic structure and the thermodynamic properties of molecular fluids in bulk, interfacial, or confined conditions at a much more modest computational cost than  molecular-dynamics or Monte-Carlo simulations. 
A current challenge concerns their implementation in three dimensions in order to describe molecular liquids, solutions, and mixtures in  complex environments such as atomistically-resolved solid interfaces or biomolecular media. There have been a number of recent efforts in that direction using 3D-RISM\cite{Beglov-Roux97,kovalenko-hirata98,red-book,yoshida09,kloss08-jcp,kloss08-jpcb}, molecular density functional theory \cite{ramirez02,ramirez05,ramirez05-CP,gendre09,zhao11,borgis12,zhao-wu11,zhao-wu11-correction}, lattice field\cite{azuara06,azuara08} or Gaussian field\cite{ varilly11} theories.

For condensed homogeneous and inhomogeneous fluids, the hard-sphere (HS) model plays a central role. It provides not only a good physical representation of colloidal dispersions where the range of inter-particle attraction is typically much smaller than the particle size, but also an invaluable reference system for studying the properties of simple liquids where the structure is predominantly determined by the short-ranged repulsion. 
In this respect, following the precusor work of Percus for one dimensional systems\cite{percus76,vanderlick89} and weighted density ideas by Tarazona and Evans\cite{tarazona-evans84,tarazona84}, the Rosenfeld's derivation in 1989 of a quasi-exact DFT for inhomogeneous hard-sphere mixtures in three dimensions,  the fundamental measure theory (FMT), constitutes a major advent of modern statistical mechanics.\cite{rosenfeld89} Several extensions or variants  of Rosenfeld's FMT were proposed subsequently. One year later, an  alternative,  scalar rather than vectorial formulation of FMT was derived by Kierlik and Rosinberg (KR)\cite{kierlik-rosinberg90,kierlik-rosinberg91}; this formulation (which was later shown to be mathematically equivalent to the Rosenfeld's original functional\cite{phan93}) will be the focus of the present work.  On the other hand, it was soon realized that FMT in its original form  was able to describe very accurately fluids  at interfaces but showed serious limitations for the description of the solid phase and liquid-solid transition, or  that of  highly confined fluids. Several successful solutions were proposed in the following two decades to extend FMT to the solid\cite{ohnesorge93,rosenfeld96,rosenfeld97,groh-mulder00,groh00,tarazona00,tarazona02,cuesta02,lutsko06}, including a tensorial correction to the vectorial formulation that  is able to  satisfy various  dimensional crossover requirements\cite{tarazona00,tarazona02,cuesta02} . Besides, an extension of the Rosenfeld's  vectorial version, based on the Mansoori-Carnahan-Starling-Leland (MCSL) hard-sphere equation of state rather than the Percus-Yevick (PY) one, was proposed  independently by Roth et al.\cite{roth02} (the so-called White-Bear (WB) version) and Wu et al. \cite{yu02} (modified fundamental measure theory, MFMT). The WB version was made compatible with the Tarazona's tensorial extensions to describe crystalline phases\cite{roth02}.  For  recent reviews of FMT and of the twenty years of efforts that have followed to improve on the initial Rosenfeld's proposal, see Refs\cite{Wu09, roth-review10}.

The existence of a hard-sphere reference for the DFT of classical fluids has promoted recently a great deal of applications of this approach to the study of atomic-like and polymeric fluids\cite{wu02,jin10}, as well as simplified models of aqueous solutions\cite{oleksy10} or ionic liquids\cite{jiang11,jiang-cpl11}, both  at interfaces or in confinement. 
Most of those studies are limited however to  planar, cylindrical or spherical symmetries. To date, there have been few applications of classical DFT in three dimensions, in order to cope with fluids in complex molecular environments\cite{ramirez02,zhao11,gillespie10}. In particular only  a few 3D implementation of FMT have been described in the literature\cite{frink00,sears03,gillespie10}. All of them are based on the original Rosenfeld's vectorial formulation. This formulation involves four scalar weigthed densities and two vectorial ones, making a total of ten weighted density components to be handled. In this paper, we propose the first  3D   implementation  of the Kierlik and Rosinberg scalar version of FMT (KR-FMT), using  the Percus-Yevick  or Carnahan-Starling variants that are both described in the original KR paper. This formulation involves only four scalar weighted densities which, as will be seen, leads to a substantial computer speedup for three-dimensional applications with respect to the vectorial version, especially for multicomponent systems. 

The needs for a three-dimensional implementations of FMT are broad and go beyond the hard-sphere model per se. Obviously, the hard-sphere functional can be used as a reference for representing  the hard core interactions in a molecular system, and other interactions such as Lennard-Jones or Yukawa attraction and  Coulombic interactions can be added in the functional  as a mean-field perturbation in order to model realistic molecular fluids, including water and ionic solutions\cite{oleksy10}. Such functionals deserve further developments and  applications to the case of complex molecular environments. Our goal is here another. We have constructed recently a three-dimensional molecular density functional theory (MDFT) approach to solvation in molecular liquids that is based on the knowledge of the angular-dependent direct correlation function of the pure solvent (the so-called homogeneous reference fluid approximation, HRF\cite{ramirez02}). This approach amounts to a second-order Taylor expansion of the free-energy with respect to the density  around the homogeneous fluid reference. It is connected to the hypernetted chain approximation (HNC) in integral equation approaches and amounts to  incorporate in the functional only the  homogeneous-solvent two-body correlations. Such approximation proved to be accurate for polar molecular fluids such as acetonitrile\cite{gendre09,zhao11,borgis12}, but appeared clearly insufficient in the case of water. In Ref.~\cite{zhao11}, we proposed to introduce empirical three-body correction terms inspired by the Molinero's water model\cite{molinero09} that re-introduce some missing tetrahedral symmetry in the HRF functional. Another possible many-body corrections are those induced by the hard-core interactions and that can be inferred from the hard-sphere FMT functional. Such approximation is at the heart of the reference HNC  approximation (RHNC) in integral equation theories\cite{lado73} and was pushed by Rosenfeld for DFT\cite{rosenfeld93}.  It was invoked recently by Zhao et al in a post-treatment of ionic microscopic solvation profiles by DFT in order to estimate the solvation free-energies\cite{zhao-wu11,zhao-wu11-correction}. This is thus the type of corrections that we investigate here. We limit ourselves in a first step to the solvation of hydrophobic solutes which, besides yielding simpler functional forms (the solvent angular dependence may be omitted), deserves special studies since hydrophobic solvation has been recently at the center of many debates\cite{lum99,chandler02,chandler-varilly11,giovambattista08,giovambattista09,matysiak11}. Let us state from the beginning that we will be dealing in this paper with the hydration of microscopic solutes\cite{pratt-chandler77} and that, at the present stage,  macroscopic effects such as dewetting are not intended to be contained in the functional.  

This paper is organized as follows. Section 2 recalls the basic principles of  hard-sphere fundamental measure theory in the original Rosenfeld's vectorial formulation and in the Kierlik-Rosinberg scalar version. Section 3 discusses the relative  practical merits of the two formulations, describes the implementation of the KR functional in three dimensions, as well as a few tests of the method. In section 4, the FMT 3D-implementation is used to add N-body hard-sphere corrections to a density functional for water in order to describe the solvation of microscopic hydrophobic solutes.

\section{Fundamental Measure Theory: Scalar versus Vectorial Formulation}

 \label{sec:FMT}

We consider a model fluid mixture composed of $N_s$  species represented by hard spheres of radius $R_i$ and bulk density $\rho_i^0$. The fluid is subjected to an external perturbation, for example a solid interface or a molecular solute of arbitrary shape embedded in the fluid, that creates for each species $i$ a position-dependent external potential $V_i(\rr)$ and thus an inhomogeneous density $\rho_i(\rr)$. The grand potential of the perturbed system can be expressed as a functional of the inhomogeneous densities and can be evaluated relatively to the homogeneous fluid 
\be
 \label{eq:F_definition}
   \Omega[\{\rho_i(\rr)\}]  = \F[\rhois] +  \Omega[\rhoiOs]
\ee
Following the general scheme of classical density functional theory\cite{evans79,evans92}, the functional $\F(\rhois)$ can be decomposed 
into an ideal, an external and an excess contribution, according to
 \bea
 \label{eq:F_components}
\F[\rhois]  &=& \F_{id}[\rhois]  + \F_{ext}[\rhois]  \nn \\
& +&  \F_{exc}[\rhois  - \F_{exc}[\rhoiOs] \nn \\
& -& \sum_i \mu_{exc}^i  \int d\rr \left( \rho_i(\rr) - \rho_i^0 \right)
\eea
where
\bea
\label{eq:Fid}
\F_{id}[\rhois] &= & k_BT \sum_i \int d\rr \,  \rho_{i}(\rr) \ln\left(\frac{\rho_{i}(\rr)}{\rho_i^0}\right) \nn \\
& &- \rho_{i}(\rr) + \rho_i^0  \\
\label{eq:Fext}
\F_{ext}[\rhois] &= &\sum_i \int d\rr \, V_i(\rr) \rho_i(\rr)
\eea
with $k_B$ is the Boltzmann constant and  $T$ is the temperature. $\F_{exc}(\rhois)$ is the excess functional for the hard-sphere fluid and $\mu_{exc}^i$  is the bulk excess chemical potential of each species  defined by
\be
\mu_{exc}^i = \frac{\delta \F_{exc}[\rhois]}{\delta \rhoi} |_{\rhois = \rhoiOs} 
\ee
In the fundamental measure theory introduced by Rosenfeld\cite{rosenfeld89}, the excess functional for the hard-sphere fluid can be written in terms of a set of $N_w$ weighted densities, ${n_\alpha(\rr)}$: 
\be
\label{eq:Fexc}
\F_{exc}[\rhois] = k_BT \int d\rr \, \Phi(\{n_\alpha(\rr)\})
\ee
with 
\be
\label{eq:weighted_densities}
n_\alpha(\rr) = \sum_i \int d\rr' \, \rho_i(\rr') \, \omega_\alpha^i(\rr - \rr') = \sum_i \rho_i(\rr) \star \omega_\alpha^i(\rr )
\ee
where $\omega_\alpha^i(\rr)$ are geometrical weight functions to be defined below and $\star$ indicates the convolution of the microscopic densities by those weight functions. 
The functional derivative of this excess free energy with respect to the densities  is given by:
\begin{eqnarray}
\dfrac{\delta F_{exc}}{\delta \rho_i (\mathbf{r})} & = & k_BT \sum_{\alpha}\int d\rr' \, \dfrac{\partial\Phi}{\partial n_{\alpha}(\mathbf{r'})} \dfrac{\partial n_{\alpha}(\mathbf{r'})}{\partial \rho_i(\mathbf{r})} \nn \\
& = & k_B T  \sum_{\alpha} \int d\rr' \, \dfrac{\partial\Phi}{\partial n_{\alpha} (\mathbf{r'})} \, \omega_i^{(\alpha)}(\mathbf{r}-\mathbf{r'}) \nn \\
\label{eq:dFexc}
& = & k_B T \sum_{\alpha} \dfrac{\partial\Phi}{\partial n_{\alpha} (\mathbf{r})} \star \omega_i^{(\alpha)}(\mathbf{r}),
\end{eqnarray}
which appears as  the convolution of the partial derivatives of $\Phi$ by the weight functions. The equilibrium inhomogeneous densities in the presence of the external potential $V_i(\rr)$ are obtained by minimization of the functional defined above, which is equivalent to solving the following Euler-Lagrange equation for all of the species
\be
\label{eq:Euler-Lagrange}
\dfrac{\delta F}{\delta \rho_i (\mathbf{r})} = k_BT \ln\left(\frac{\rho_{i}(\rr)}{\rho_i^0}\right) + V_i(\rr) + \dfrac{\delta F_{exc}}{\delta \rho_i (\mathbf{r})} - \mu_{exc}^i = 0
\ee

In the original Rosenfeld's derivation there are four scalar weight functions, $\omega_\alpha^i(\rr), \alpha = 0, 1, 2, 3$, and two vectorial ones $\vec{\omega}_1(\rr), \vec{\omega}_2(\rr)$ per species $i$ that are defined by
\bea
\label{eq:w3}
\omega_3^i(\rr) &= &\Theta(R_i - r) \\
\label{eq:w2}
\omega_2^i(\rr) &=& 4 \pi R_i \, \omega_1^i(\rr) = 4 \pi R_i^2 \, \omega_0^i(\rr)  =\delta(R_i - r) \\
\label{eq:wvec2}
\vec{\omega}_2^i(\rr) &=& 4 \pi R_i \, \vec{\omega}_1^i(\rr) = \frac{\rr}{r} \, \delta(R_i - r) 
\eea
$\Theta(r)$ denotes the Heaviside function and 
$\delta(r)$ the Dirac distribution.
The excess free-energy density $\Phi$  derived by  Rosenfeld for Eq. \ref{eq:Fexc} is a function of  the three position-dependent weighted densities, $n_\alpha(\rr), \alpha=0, 1, 2, 3$,  and of the two vectorial ones, $\vec{n}_1(\rr), \vec{n}_2(\rr)$, which generates  in the homogeneous limit the Percus-Yevick equation of state for hard-sphere mixtures. Starting from the generalization of the Carnahan-Starling (CS) equation of state to mixtures (namely the Mansoori-Carnahan-Starling-Leland  equation (MCSL))   instead of PY, Roth et al\cite{roth02} and Wu et al \cite{wu02} were later able to obtain a modified expression  based on the same definition of the weighted densities (either called white-bear (WB) version or modified FMT version (MFMT)). This modified version of FMT takes advantage of the fact that the CS expression provides one a better equation of state that PY.

Ten years before those latest developments, Kierlik and Rosinberg  were able to derive an alternative version of FMT  which involves only four scalar weight functions $\omega_\alpha^i(\rr), \alpha = 0, 1, 2, 3$.\cite{kierlik-rosinberg90,kierlik-rosinberg91}. The last two weights are identical to Eq.~\ref{eq:w3}-\ref{eq:w2}, whereas the first two ones are given by
\bea
\label{eq:w1}
\omega_1^i(\rr)& = &\frac{1}{8\pi} \delta'(R_i - r) \\
\label{eq:w0}
\omega_0^i(\rr) & = & \frac{1}{8\pi} \, \delta''(R_i - r) + \frac{1}{2\pi r} \, \delta'(R_i - r) 
\eea
Those weight functions appear naturally in the derivation as the inverse Fourier transforms of
\begin{eqnarray}
\omega_3^i(k)&=&\frac{4\pi}{k^3}(\sin(kR_i)-kR_i\cos(kR_i)) \nn \\
\omega_2^i(k)&=&\frac{4\pi R_i}{k} \sin(kR_i) \nn \\
\omega_1^i(k)&=&\frac{1}{2k}(\sin(kR_i)+kR_i\cos(kR_i)) 
\label{eq:wk}\\
\omega_0^i(k)&=&\cos(kR_i)+\frac{kR_i}{2}\sin(kR_i) \nn
\end{eqnarray}
Although the main part of the papers by Kierlik and Rosinberg relies on  a PY expression for the excess free energy density
\begin{equation}
\label{eq:Phi_PY}
\Phi^{\text{PY}}[n_\alpha]=-n_0\ln(1-n_3)+\frac{n_1n_2}{1-n_3}+\frac{1}{24\pi}\frac{n_2^3}{(1-n_3)^2},
\end{equation}
the authors do mention in their conclusion that a CS (more precisely  MCSL) expression could be used instead
\begin{eqnarray}
\label{eq:Phi_CS}
\Phi^{\text{CS}}[n_\alpha]&=&\left(\frac{1}{36\pi}\frac{n_2^3}{n_3^2}-n_0\right)\ln(1-n_3)\nn \\
& &+\frac{n_1 n_2}{1-n_3}+\frac{1}{36\pi}\frac{n_2^3}{(1-n_3)^2n_3}.
\end{eqnarray}
They point out  the fact that this expression is more precise than the PY one, but  using it while keeping the expression of the weights unchanged leads to thermodynamic inconsistencies; those inconsistencies are indeed present in the WB or MFMT formulations too. There is clearly a trade off to be made between precision and theoretical consistency. It was later shown by Phan et al. that the  Kierlik and Rosinberg's approach is mathematically equivalent to the original vectorial version.\cite{phan_equivalence_1993}  On a practical point of view, however, and especially in the perspective of  3D applications, the KR formulation is advantageous with respect to the Rosenfeld's formulation  since the number of weighted densities is reduced. So is the number of convolutions and thus the number of 3D-Fast Fourier Transforms (3D-FFT) to be performed. This technical point is discussed below in more details. Before proceeding, we note again that the functionals considered above  are well suited to describe inhomogeneous liquids at interfaces or in loose confinement, but they are known to fail for crystalline phases or highly confined conditions. In those cases, various extensions of 3D-FMT have been devised\cite{rosenfeld96,rosenfeld97,tarazona00,tarazona02,cuesta02}.  They lie outside the scope of the present study.

\section{Implementation of the Kierlik-Rosinberg Functional in 3D} 

As mentioned in the introduction, there have been a few 3D-implementations of FMT proposed in the literature, the first one by Frink and Salinger\cite{frink00}. All of them are based on the Rosenfeld's vectorial formulation. Whatever the formulation chosen, however, a natural way to solve  the FMT equations  in 3D is to discretize them on a 3D orthorombic grid and to handle the convolutions through 3D-FFT's. A typical minimization algorithm  requires one to provide at each minimization step the value of the functional and of the functional derivatives for a given set of densities $\{\rho_i(\rr)\}$. If we denote by $N_w$ the number of weight functions to be considered, the numerical procedure  involves  1)  to transform the densities in Fourier space to $\{\rho_i(\kk)\}$,  2)  to multiply those densities by the $N_w \times N_s$ weight functions and sum the products  over the different species to get the weighted densities (Eq.~\ref{eq:weighted_densities}), 3) to transform back the $N_w$ weighted densities to real space to compute the excess free energies (eq.~\ref{eq:Fexc}) and the partial derivatives with respect to those weighted densities $\frac{\partial \Phi}{\partial n_\alpha}(\rr)$, 4) to transform those quantities  to k-space and, for each species, to multiply them by the weight functions and sum, and finally 5) to  back transform the results to real space to get $\dfrac{\delta F_{exc}}{\delta \rho_i (\rr)}$ for all of the species (Eq.~\ref{eq:dFexc}).

The whole procedure sums up to $2(N_s + N_w)$ FFT's to be performed.  $N_w=4$ for the KR scheme whereas $N_w=10$ in the vectorial formulation (4 scalar weights + ($2 \times 3$) vectorial components in 3D).  For a one component system, one can take advantage in the vectorial formulation of the relationships that exist between the weights  and reduce the number of independent weights to be considered to $N_w=5$. In this case the speedup of the scalar versus the vectorial formulation appears rather  marginal: each cycle requires 10 forward and backward FFT's instead of 12, thus a  $\sim20 \%$ difference. The reduction of the number of independent weights does not apply to multi-component mixtures. The balance thus becomes 12 FFT's versus 24 for a two-component system and  $(8 + 2N_s)$  versus $(20 + 2N_s)$ in the general case. The expected speedup is thus, in this more generic case, of $100 \%$ and more.

For those reasons, we propose in this paper the first 3D-implementation  of the Kierlik-Rosinberg's version of FMT.  We have implemented both the PY and CS (MCSL) versions, which only differ in the expression of the excess free-energy density, Eq.~\ref{eq:Phi_PY} or~\ref{eq:Phi_CS}, and the corresponding partial derivatives with respect to the weighted densities. For an arbitrary number of species in the HS mixture, the densities are  discretized on an orthorombic grid of dimension $L_x \times L_y \times L_z$. The external potentials $V_i(\rr)$ for every species are first pre-computed and tabulated. This potential might originate from hard  walls, or from  molecular solutes embedded in the mixture and  described in terms of site-distributed hard-sphere repulsions or Lennard-Jones  interactions. For a given external potential, the FMT functional described by Eq.~\ref{eq:F_components}-\ref{eq:weighted_densities} is then minimized using the forward-backward FFT scheme described above. The minimization is performed in direct space with respect to  the fictitious "wave-functions" $\psi_i(\rr)$, defined by 
$\rhoi = \psi_i(\rr)^2$, in order to avoid spurious negative values of the densities that would make the logarithm term of the ideal part of the free energy functional diverge. As a minimization algorithm, we had recourse to  the L-BFGS quasi-Newton optimization routine\cite{BFGS} which is optimized  to handle very large systems  and requires one, at each step, to provide free-energy value and gradients. The gradients are known analytically as in 
eq.~\ref{eq:Euler-Lagrange}.

\begin{figure}
  \includegraphics[width=1.05\columnwidth]{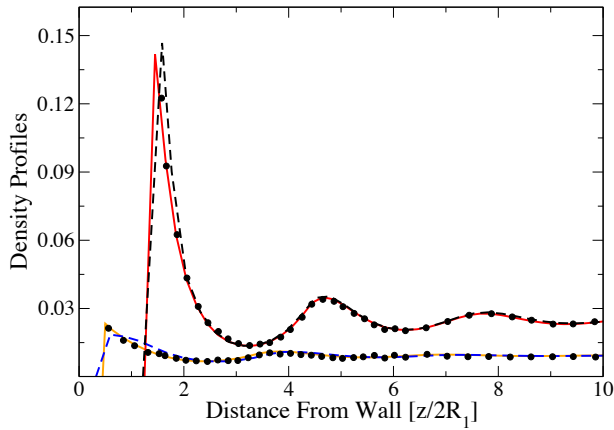}
\caption{Reduced density profiles $8R_1^3\rho(z)$ versus the distance from the wall $z/2R_1$ of a binary hard-sphere mixture near a hard wall at diameter ratio $R_2/R_1=3$ and bulk reduced densities $0.0260$ and $0.0104$. The lines represent the 3D-FMT results using 3  or 6 grid points per hard-sphere diameter in the z-direction (dashed and solid lines, respectively). The black dots are the Monte-Carlo reference simulation data from Tan et al.~\cite{tan_hardsphere_1989}.
\label{fig:mixtures_density_profiles}}
\end{figure}

\begin{figure}
  \includegraphics[width=1.1\columnwidth]{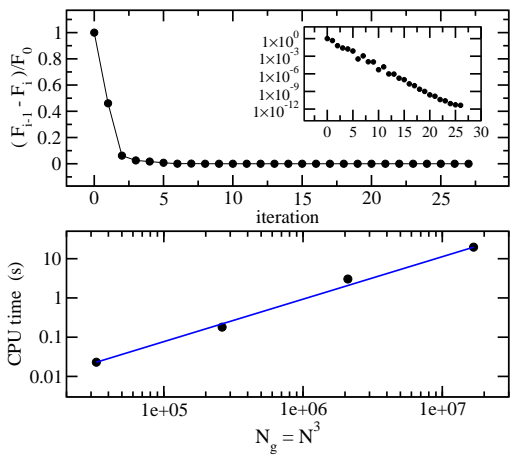}
\caption{Top: Typical plot of the free energy difference between two successive steps (normalized by the initial energy) versus L-BFGS minimization-step number (Here a benzene molecule in a one-component HS reference fluid modeling SPC water; see Fig.~\ref{fig:benzene_perspective}). The inlet represents the same  with a logarithmic scale in ordinates. Bottom: CPU time per minimization step versus number of 3D-grid points. The circle correspond in increasing order to N= 32, 64, 128, and 256. 
\label{fig:convergence}}
\end{figure}

We first illustrate our FMT implementation for multi-component mixtures for  the classical test case of a two component hard-sphere fluid with radii $R_1$ and $R_2=3R_1$ near a hard wall\cite{rosenfeld93,sears03}, for which  reference Monte-Carlo calculations are available in the literature \cite{tan_hardsphere_1989}. We show in  Fig~\ref{fig:mixtures_density_profiles} that the code converges and  gives sensible results for even a  low grid resolution  of 3 points per small hard-sphere diameter, $\sigma_1 = 2 R_1$, in all directions. The 3D KR--FMT results appear already in excellent agreement with the simulation data when using a finer resolution of 6 points/$\sigma_1$  in the z-direction while leaving the xy-resolution unchanged.
A convergence criterium of $10^{-6}$ is reached after 10 minimization iterations starting a a uniform bulk mixture. Such typical convergence is further illustrated in Fig.~\ref{fig:convergence} for  the solvation of  a neutral benzene molecule, represented by 12 lennard-Jones atomic sites  in a one-component Lennard-Jones fluid modeled by a hard-sphere FMT functional with radius $R=1.25 \, \AA$ and  at a liquid density 
$\rho_0 = 0.03328$ particles/$\AA^3$ (a simplified representation of water at ambient conditions, see the next section). Starting with a guess density $\rho(\rr)=\rho_0 \, \exp(-\beta V(\rr))$, where $V(\rr)$ is the external potential, it is seen that the convergence is basically exponential as a function of the minimization step and that typically 10-15 iterations are required to get fully converged results. We show in Fig.~\ref{fig:convergence}  that the required computer time grows linearly with the number $N_g$ of grid points. 
Performed on the single processor of a standard laptop or desktop computer, a full minimization cycle requires  between a few seconds  for $N_g=64^3$ and a few minutes for $N_g=256^3$. For a solvent of the size of water ($\sigma \simeq 3.0 \AA$), a typical resolution of 3-4 points/$\AA$ is sufficient to get accurate free-energies and densities. With such resolution, one can foresee a possible application of the method to rather large molecular system, requiring  box sizes up to 100 $\AA$.

\section{Application to hydrophobic solvation}

In continuation to previous works on molecular density functional theory\cite{ramirez02, ramirez05,ramirez05-CP,gendre09,zhao11,borgis12,zhao-wu11,zhao-wu11-correction}, we apply our  implementation of KR-FMT to improve the MDFT description of molecular solutes in liquid water at ambient conditions (bulk density 
$\rho_0 = 0.03328$ molecules/$\AA^3$).  In MDFT, the solvent response to the solute external field  is described in terms of a functional of the inhomogeneous position and orientation density, $\rho(\rr, \Omega)$. If the solute is modeled as a purely hydrophobic entity, bearing no electrostatic multipole, and if the solvent position-orientation coupling is neglected (which is a good approximation for water), the angular dependence can be omitted, and the functional can be expressed in terms of the isotropic number density, 
$\rho(\rr) = \int d\Omega \, \rho(\rr, \Omega)$:
\bea
\F[\rhow] &= & k_BT \int d\rr \, \left[ \rho(\rr) \ln\left(\frac{\rho(\rr)}{\rho_0}\right) - \rho(\rr) + \rho_0 \right] \nn \\
& &+  \int d\rr \, V(\rr) \rho(\rr) \nn \\
& &-  \frac{k_BT}{2}  \int d\rr d\rr' \Delta\rho(\rr) \, \Delta\rho(\rr') \, c_S(|\rr - \rr'|;\rho_0) \nn \\
& &  + \F_B[\rhow]
\label{eq:Fwater}
\eea
with $\Delta\rho(\rr) = \rho(\rr) - \rho_0$. $V(\rr)$ is the external potential to be defined below. The last two terms represent the excess free energy, $\F_{exc}[\rhow]$, which is decomposed into a homogeneous reference fluid (HRF) term, involving the isotropic direct correlation function of the pure solvent at the density $\rho_0$, $c_S(r;\rho_0)$, and a correction term or "bridge" term (in reference to integral equation theory), which is basically unknown, but can be  formally expanded in terms of the pure solvent three-body,..., N-body  direct correlation functions. The approximation that we propose here is rather standard\cite{rosenfeld93,biben98,zhao-wu11} and consists in replacing the unknown bridge for liquid water by the exact bridge  of an equivalent hard-sphere fluid 
\bea
\F_B[\rhow] &= &\F_{exc}^{HS}[\rhow] -   \F_{exc}^{HS}[\rho_0] - \mu_{exc}^{HS} \int d\rr \Delta\rhow \nn \\
& +  & \frac{k_BT}{2} \int d\rr d\rr' \Delta\rho(\rr) \, \Delta\rho(\rr') \, c_S^{HS}(|\rr - \rr'|;\rho_0)  
\label{eq:FBwater}
\eea
The first three terms represent the one-component hard-sphere KR-FMT excess functional defined in the previous section and  the associated chemical potential yielding equilibrium at $\rhow =  \rho_0$. The fourth term involves the  direct correlation function of the HS fluid at the same density, i.e 
\be
c_S^{HS}(|\rr - \rr'|;\rho_0) = - \frac{\delta^2 \F_{exc}^{HS}[\rho]}{\delta\rho(\rr) \delta\rho(\rr')}|_{\rhow = \rho_0} .
\ee
This function can be easily obtained in Fourier space as
\be
c_S^{HS}(k;\rho_0) = - \sum_{\alpha,\beta} \frac{\partial^2\Phi}{\partial n_\alpha \partial n_\beta}(\{ n_\gamma^0 \}) \, \omega_\alpha(k) \omega_\beta(k)
\label{eq:cS_HS}
\ee
where $\{n_\gamma^0\}$ represent the weighted densities for a uniform fluid  of density $\rho_0$ and the 
$\omega_{\alpha,\beta}(k)$ are the weights of eq.~\ref{eq:wk}. The second derivatives have to be taken for the PY or CS functions of eqs~\ref{eq:Phi_PY} or \ref{eq:Phi_CS}; the corresponding functions are reported in the Appendix. Note that defined as in eq.~\ref{eq:FBwater}, $\F_B[\rhow]$   carries an expansion in $\Delta \rho$  of order 3 and higher that corrects the second order expansion of the excess free energy in eq.~\ref{eq:Fwater}. 

\begin{figure}
  \includegraphics[width=1.05\columnwidth]{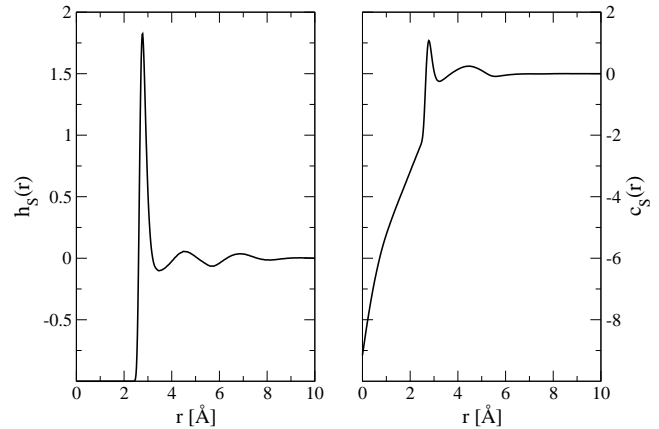}
\caption{Left: Oxygen-oxygen isotropic pair distribution function of SPC water, $h_S(r)$, computed by molecular dynamics. Right: corresponding direct correlation function, $c_S(r)$, obtained by Ornstein-Zernike inversion using the Baxter direct-space method.
\label{fig:hs+cs_spc}}
\end{figure}

In this approach, two elements should be further defined. First the direct correlation of water. In principle it can be extracted from the experimental oxygen-oxygen structure factor. Since our further comparison will be with respect to molecular dynamics simulations carried out with the SPC water model, we have computed $c_S(r;\rho_0)$ for this model. To do so we have re-generated the well-known oxygen-oxygen pair distribution function by carrying out molecular dynamics simulations with 4096 water molecules and a box size of $\sim 50 \AA$; see Fig.~\ref{fig:hs+cs_spc}. The corresponding direct correlation function can be deduced by solving the Ornstein-Zernike equation. This can be done quite naturally in Fourier-space\cite{hansen}. To avoid the numerical problems that occur in this case at small k-values,  we have used instead the direct space method of Baxter\cite{baxter70} combined with the variational method of Dixon and Hutchinson\cite{dixon77}; see Ref.~\cite{ramirez05-CP} for details. This method imposes that the direct correlation function vanishes beyond a cut-off value  that we set to $R_c=8.7 \AA$. The corresponding function $c_S(r;\rho_0)$ is plotted in Fig.~\ref{fig:hs+cs_spc}. A second necessary ingredient is to fix the radius $R$ of the equivalent hard-sphere fluid. A natural value is around $R \simeq 1.25 \AA$ that corresponds to the hard core, i.e. the region where $h_S(r) = 0$ in Fig.~\ref{fig:hs+cs_spc}. This choice can be confirmed by a striking fact noticed by Chandler and Varilly in Ref.~\cite{chandler-varilly11} (see their figure 6): the statistics of spontaneous empty cavities in SPC water, that they tightly link in their Gaussian field analysis to the mechanism of hydrophobic solvation, are quite similar to those obtained for a hard-sphere liquid at a reduced density  $\rho^* = 8 \rho_0 R^3 = 0.5$, yielding $R \simeq 1.25 \AA$ at the water density. Small variations around that value can be conceived and for reasons described below we were led to choose $R=1.27 \AA$.

With the previous elements in hand, the functional of eqs~\ref{eq:Fwater}-\ref{eq:FBwater} can be minimized in the  external Lennard-Jones potential field imposed by a molecular solute placed at the center of a cubic
box. It is defined as 
\be
V(\rr) = \sum_i 4 \epsilon_{wi} \left[ \left( \frac{\sigma_{wi}}{|\rr - \rr_i |} \right)^{12} - 
\left( \frac{\sigma_{wi}}{|\rr - \rr_i |} \right]^{6} \right]
\ee
where the $\rr_i$'s stand for  the  positions of the solute atomic sites and $\sigma_{wi}, \epsilon_{wi}$  are the site-water Lennard-Jones parameters (using Lorentz-Berthelot mixing rules and the SPC parameters for water).

\begin{figure}
   \includegraphics[width=0.9\columnwidth]{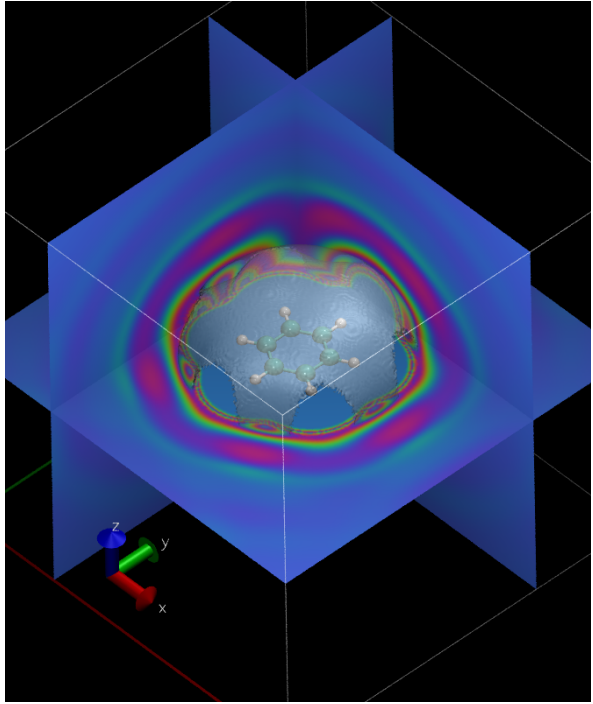}
\caption{Three-dimensional representation of the reduced density of SPC water around a benzene molecule obtained by minimization  of the functional. Blue to red indicate low to high densities up to $\rho(\rr)/\rho_0 \simeq 3.5$. The transparent grey surface that appears above the molecule represents the isosurface $\rho(\rr)/\rho_0=2.0$. 
\label{fig:benzene_perspective}
}
\end{figure}

\begin{figure}
  \includegraphics[width=1.05\columnwidth]{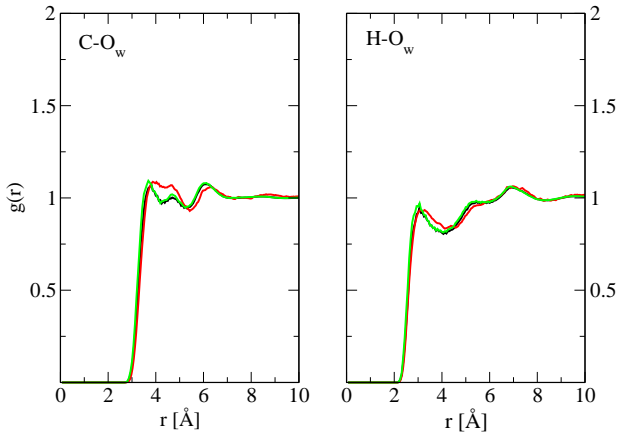}
\caption{Site-oxygen radial distribution functions for a benzene molecule in SPC water: MD results (red line) compared to the DFT results with and without the HS bridge term with $R=1.27 \AA$ (black and green lines, respectively). 
\label{fig:gr_benzene}
}
\end{figure}

\begin{figure}
  \includegraphics[width=0.5\columnwidth]{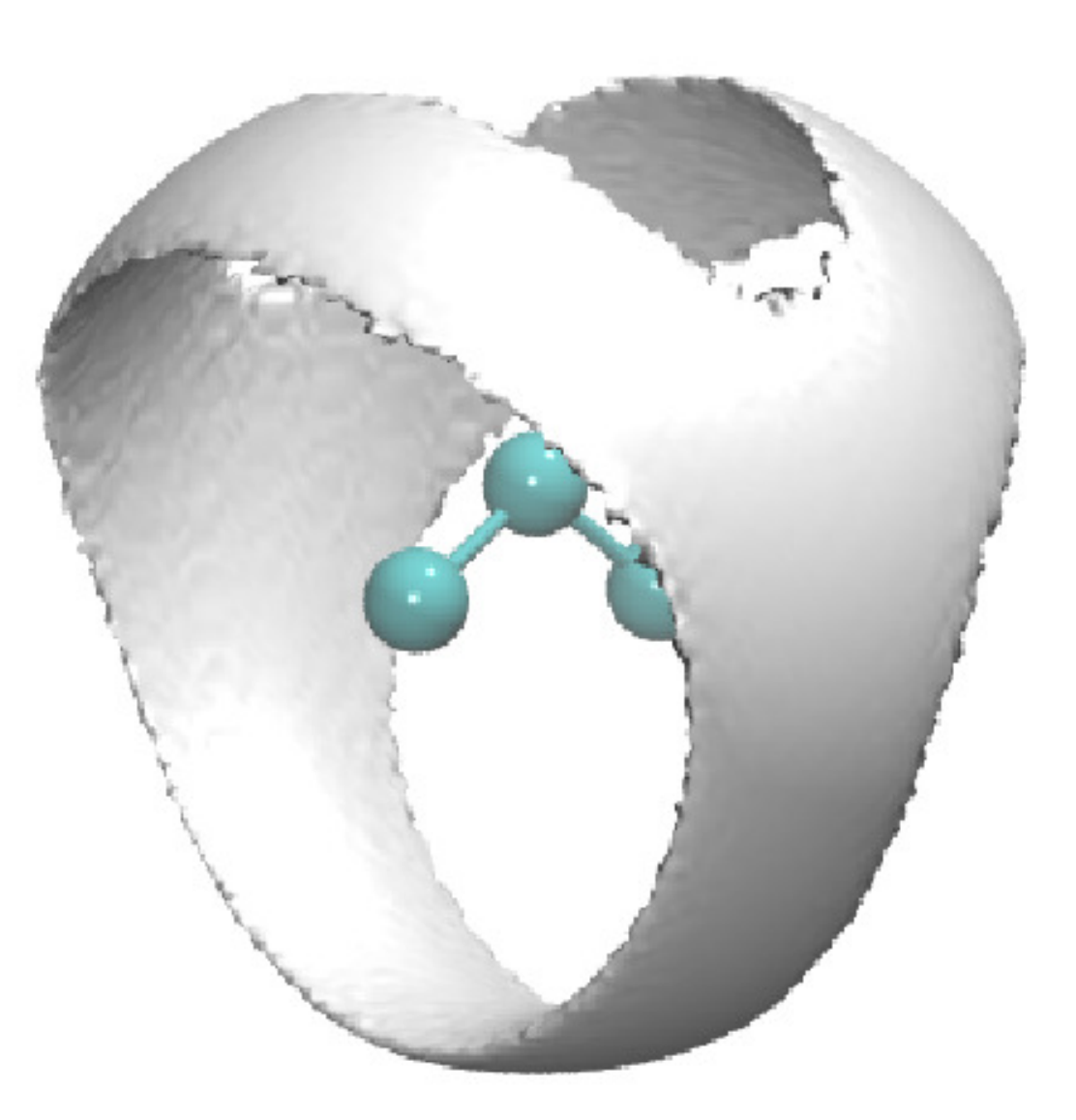}
  \includegraphics[width=1.05\columnwidth]{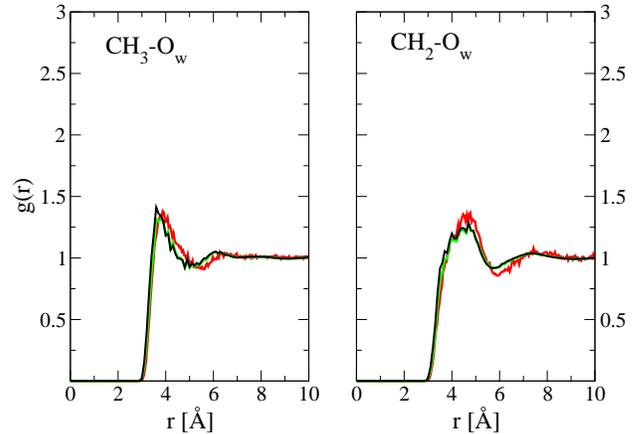}
\caption{Site-oxygen radial distribution functions for a propane molecule in SPC water: MD results (red line) compared to the DFT results with and without the HS bridge term with $R=1.27 \AA$ (black and green lines, respectively). The top figure is a representation of the molecule and of  the solvent reduced density isosurface corresponding to $\rho(\rr)/\rho_0=2.0$.
\label{fig:gr_propane} 
}
\end{figure}

For the the KR-FMT excess free-energy terms, we have used either the PY or CS version, eqs~\ref{eq:Phi_PY} and ~\ref{eq:Phi_CS} (with very little influence of this choice on the results, as will be seen). In addition to those terms that can be  handled as described in the previous section, one needs  to compute the $c_S$ and $c_S^{HS}$ quadratic terms; since they appear as convolutions, this is easily done by forward-backward Fourier transform as in a regular HRF approximation\cite{ramirez02}. The procedure is illustrated in Fig.~\ref{fig:benzene_perspective}, representing the three-dimensional density obtained by minimization around a benzene molecule (Lennard-Jones parameters from Ref.~\cite{laaksonen98}, no electrostatics). In Fig.~\ref{fig:gr_benzene}, we compare the molecule site-water oxygen (C-O$_w$ and H-O$_w$) pair distribution functions obtained by DFT with and without the bridge term of eq.~\ref{eq:FBwater} to the same quantities  generated by molecular dynamics simulations (one solute and 512 SPC water molecules). Two features should be noted. First  the 
bridge term turns out to have little influence on the overall microscopic structure, although it has a noticeable influence on the computed solvation free energies (see below).  Secondly,  the agreement to MD can be qualified as quite satisfactory, despite a disagreement in the shape of the first peak for C-O$_w$. The same type of comparison is drawn
in Fig.~\ref{fig:gr_propane} for propane. The geometry and parameters of Ashbaugh et al.\cite{ashbaugh_hydration_1998} were used with a unified description of the CH$_2$ and CH$_3$ groups. Again the agreement with the MD results is quite good, with a slight underestimation of the first peak width for CH$_3$-O$_w$. The influence of the bridge term on the structure remains marginal.

\begin{figure}
  \includegraphics[width=1.05\columnwidth]{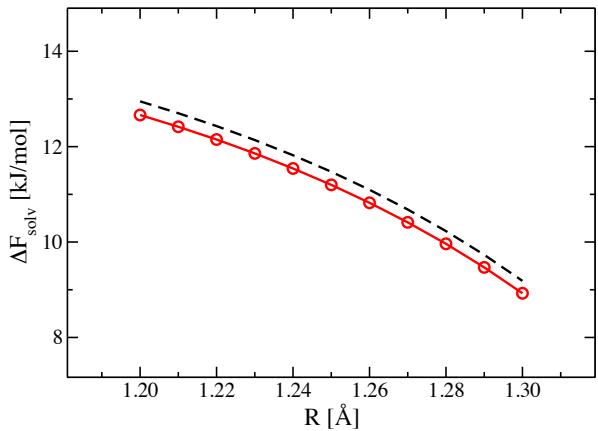}
\caption{Hydration free energy versus reference hard-sphere radius for methane, using either the Percus-Yevick (red circles and line) 
or Carnahan-Starling (black dashed line) versions of the FMT functional for the bridge term (eq.~\ref{eq:FBwater}).
\label{fig:hsolv_wrt_radius}}
\end{figure}

\begin{figure}
  \includegraphics[width=1.05\columnwidth]{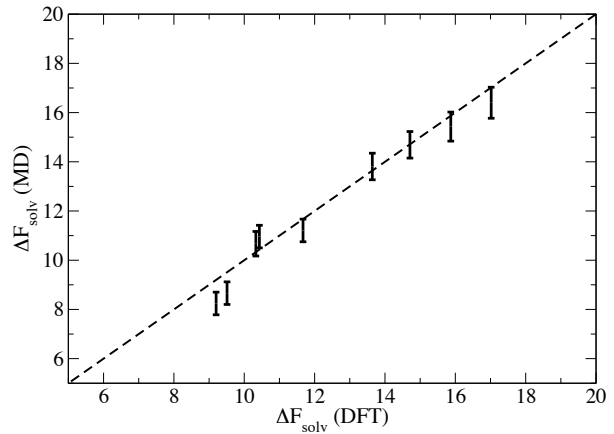}
\caption{Comparison of the free energy of hydration for rare gases and alcanes (in trans conformation), calculated either by DFT with a hard-sphere bridge function of radius $1.27 \AA$ or by MD simulations (Refs~\cite{ashbaugh_hydration_1998} and \cite{guillot_guissani_1993}). From left to right: Kr, Ar, ethane, methane, propane, Ne, butane, pentane, hexane. The vertical bars correspond to the MD error bars indicated in Table 1.
Units are kJ/mol.
\label{fig:gsolv}}
\end{figure}

Finally Table 1 compares the solvation free energies obtained for a series of rare-gas atoms  and alcanes molecules to the MD results reported by Guillot and Guissani using a particle insertion method\cite{guillot_guissani_1993} or Ashbaugh et al. using thermodynamic integration techniques\cite{ashbaugh_hydration_1998}. It is seen that the straight application  of the  HRF approximation, $\F_B[\rho]=0$,  systematically overestimate the solvation free-energies.  In Fig.~\ref{fig:hsolv_wrt_radius} we display the free-energy energy of methane obtained when adding the hard-sphere bridge term of eq.~\ref{eq:FBwater} and varying the hard-sphere radius  around $1.25 \AA$. It can be observed that the computed free-energy decreases steadily with increasing $R$ (whereas the microscopic water structure remains unaffected by the added bridge term in this parameter range, see Figs~\ref{fig:gr_benzene} and ~\ref{fig:gr_propane}). Furthermore we show that using either the CS or PY version of  the HS functional   has  a   very small  --although measurable-- effect on the results.  Retaining  the CS version, we find  that the MD value for methane in Table 1 is reached when $R \simeq 1.27 \AA$.  Keeping that value, we find a close correlation to the MD results  for the whole series of molecules. These encouraging findings are listed in Table 1 and further depicted in Fig.~\ref{fig:gsolv}. Note that for each point in the figure, the computational effort to get the solvation free energy is orders of magnitude lower  for 3D-DFT than for MD.

\begin{table}
\begin{tabular}{|c|c|c|c|}
\hline 
Molecule & MD &  DFT/HRF  &  DFT/HRF+bridge \\
\hline 
\hline 
methane & 10.96$\pm 0.46$  & 16.03 & 10.43\\
\hline 
ethane & 10.75$\pm 0.50$ & 18.50 & 10.33\\
\hline 
propane & 13.81$\pm 0.54$ & 24.86 & 13.64\\
\hline 
butane & 14.69$\pm 0.54$  & 28.56 & 14.71\\
\hline 
pentane & 15.43$\pm 0.59$  & 32.23 & 15.87\\
\hline 
hexane & 16.40$\pm 0.63$ & 35.99 & 17.02\\
\hline 
Ne & 11.21$\pm 0.46$  & 14.47 & 11.67\\
\hline 
Ar & 8.661$\pm 0.46$  & 14.04 & 9.51\\
\hline 
Kr & 8.242$\pm 0.46$ & 14.68 & 9.20\\
\hline 
\end{tabular}

\caption{Solvation free energies of rare-gas atoms and alcane molecules in SPC water computed by DFT using the HRF approximation (eq.~\ref{eq:Fwater} with $\F_B =0$) or the HRF + hard-sphere bridge approximation (eqs~\ref{eq:Fwater} and eq.~\ref{eq:FBwater} with $R=1.27 \AA$). They are compared to the MD values of Ref.~\cite{ashbaugh_hydration_1998} for alcanes and Ref.~\cite{guillot_guissani_1993} for rare gases (with the corresponding error bars). All values are in kJ/mol.}
\end{table}

\section{Conclusion}

This paper has presented the first three-dimensional implementation, to our knowledge, of the Kierlik-Rosinberg fundamental measure theory for hard-sphere mixtures. 
Since the free-energy is written in terms of convolutions of the microscopic density with respect to discontinuous weight functions, the FMT density functionals are
reputed to require very fine grids and very long recursive minimizations, at least for low dimensional applications. In the three-dimensional case, we showed that using discrete 3D-FFT's which are well suited to describe  the convolution of smooth functions with discontinuous (step-like or delta-like) distributions and using a more elaborated minimizer that the Picard iteration
scheme that is usually prescribed\cite{roth-review10}, the Kierlik-Rosinberg 
FMT functional can be efficiently minimized  for realistic systems with a grid resolution of only a few points per Angstrom and within at most a few tens of iterations.
Such implementation  constitutes a basis for tackling very diverse problems in physical chemistry involving fluids or solutions in the presence of atomistically-resolved interfaces or  confinement matrices.  To describe realistic interactions, dispersive and Coulombic contributions can be easily introduced as mean field perturbations to the hard-core functional\cite{biben98, oleksy10}. We are working presently on several applications in that context.

 We have used here the implementation for another purpose: try to infer N-body corrections in the molecular density functional theory description of water, a system for which the restriction to two-body correlations
(the so called homogeneous reference fluid approximation or HNC approximation in an integral equation context) was found to present shortcomings\cite{richardi98,zhao11}.
Hard-sphere corrections do seem to help  for the special but important  case of hydrophobic solvation. As stated in the introduction, however, the present approach is meant to describe hydrophobicity at microscopic length scales but not to account properly  for macroscopic phenomena  such as dewetting\cite{lum99,chandler02,varilly11}. Such limitation could be bypassed by adding coarse-grained contributions to the microscopic short-ranged functional.

 Furthermore the present effort has to be continued for the general situation of polar and charged solutes, for which electrostatic interactions and solvent angular dependence have to be included. In that case, despite  positive results  when  MDFT is used as a post-treatment of exact densities\cite{zhao-wu11,zhao-wu11-correction}, we have  some preliminary indications, and there are premises in the literature too\cite{richardi99,lombardero99,reddy03}, that life might not be so simple with self-consistent minimization, and that angular-independent bridge corrections might not be sufficient. Mixing the type of hard-sphere corrections studied here to the H-bonding three-body corrections introduced in Ref.~\cite{zhao11} might be a way out.

\begin{acknowledgments}
The authors acknowledge financial support from the Agence Nationale de la Recherche under grant ANR-09-SYSC-012. They are grateful to Jean-Pierre Hansen and Benjamin Rotenberg for fruitful discussions.
\end{acknowledgments}

\appendix
\section{A few useful formulas}

The first derivatives of the function $\Phi(\{n_\alpha \})$ in eq.~\ref{eq:Fexc} with respect to the weighted densities are required to compute the functional gradients and perform the minimization (see the Euler-Lagrange equation, eq.~\ref{eq:Euler-Lagrange}). The second derivatives make it possible to compute the hard-sphere direct correlation functions, eq.~\ref{eq:cS_HS}. They are  required too if a Newton-like minimization algorithm  such as GMRES\cite{saad86,sears03} is used instead of L-BFGS (which is quite efficient but memory demanding).

All those derivatives are listed below for the PY version of KR-FMT, eq.~\ref{eq:Phi_PY}:
\bea
\frac{\partial \Phi^{PY}}{\partial n_0} & = & -\ln(1-n_3) \nn \\
\frac{\partial \Phi^{PY}}{\partial n_1} & = & \frac{n_2}{1-n_3} \nn  \\
\frac{\partial \Phi^{PY}}{\partial n_2} & = & \frac{n_1}{1-n_3} + \frac{n_2^2}{8\pi(1-n_3)^2} \nn \\
\frac{\partial \Phi^{PY}}{\partial n_3} & = & \frac{n_0}{1-n_3}+\frac{n_1 n_2}{(1-n_3)^2}+\frac{n_2^3}{12\pi(1-n_3)^3} \nn \\
\frac{\partial^2 \Phi^{PY}}{\partial n_0\partial n_3} & = & \frac{\partial^2 \Phi^{PY}}{\partial n_1\partial n_2}  = \frac{1}{1-n_3} \\
\frac{\partial^2 \Phi^{PY}}{\partial n_1\partial n_3} & = &  \frac{n_2}{(1-n_3)^2} \nn \\
\frac{\partial^2 \Phi^{PY}}{\partial n_2^2} &=& \frac{n_2}{4\pi(1-n_3)^2} \nn \\
\frac{\partial^2 \Phi^{PY}}{\partial n_2\partial n_3} & = &  \frac{n_1}{(1-n_3)^2}+\frac{n_2^2}{4\pi(1-n_3)^3} \nn \\
\frac{\partial^2 \Phi^{PY}}{\partial n_3^2} &=& \frac{n_0}{(1-n_3)^2}+\frac{2 n_1 n_2}{(1-n_3)^3}+\frac{n_2^3}{4\pi(1-n_3)^4} \nn
\eea
and for the CS version, eq.~\ref{eq:Phi_PY}:
\bea
\frac{\partial \Phi^{CS}}{\partial n_0} & = & -\ln(1-n_3) \nn \\
\frac{\partial \Phi^{CS}}{\partial n_1} & = & \frac{n_2}{1-n_3} \nn  \\
\frac{\partial \Phi^{CS}}{\partial n_2} & = & \frac{n_1}{1-n_3} + \frac{n_2^2}{12\pi (1-n_3)^2 n_3}+\frac{n_2^2 }{12\pi n_3^2}\ln(1-n_3) \nn \\
\frac{\partial \Phi^{CS}}{\partial n_3} & = & \frac{n_0-n_2^3/(36\pi n_3^2)}{1-n_3}+\frac{n_1 n_2}{(1-n_3)^2}-\frac{n_2^3}{36\pi (1-n_3)^2 n_3^2} \nn \\
         & &  +\frac{n_2^3}{18\pi n_3(1-n_3)^3} -\frac{n_2^3}{18\pi n_3^3}\ln(1-n_3) \nn \\
\frac{\partial^2 \Phi^{CS}}{\partial n_0\partial n_3} & = & \frac{\partial^2 \Phi^{CS}}{\partial n_1\partial n_2}  = \frac{1}{1-n_3} \\
\frac{\partial^2 \Phi^{CS}}{\partial n_1\partial n_3} & = &   \frac{n_2}{(1-n_3)^2} \nn \\
\frac{\partial^2 \Phi^{CS}}{\partial n_2^2} & = & \frac{n_2}{6\pi n_3(1-n_3)^2}+\frac{n_2}{6\pi n_3^2}\ln(1-n_3) \nn \\
\frac{\partial^2 \Phi^{CS}}{\partial n_2\partial n_3} & = & 
-\Big(n_3(n_2^2(2-5n_3 + n_3^2) - 12 \pi n_1 (1-n_3)n_3^2) \nn \\
& &+ 2 n_2^2 (1-n_3)^3 \ln(1 - n_3)\Big)/(12 \pi (1-n_3)^3 n_3^3) \nn \\
\frac{\partial^2 \Phi^{CS}}{\partial n_3^2} & = & \Big(n_3 (n_2^3 (6 - 21 n_3 + 26 n_3^2 - 5 n_3^3) + 72 \pi n_1 n_2 (1 - n_3) n_3^3  \nn \\
& & + 36 \pi n_0 (1 - n_3)^2 n_3^3 ) + 6 n_2^3 (1 - n_3)^4 \ln(1 - n_3)\Big)  \nn \\
& & /(36 \pi (1 - n_3)^4 n_3^4) \nn
\eea
All the second derivatives that are not written are equal to zero. There are thus 6 non-vanishing second derivatives to be considered instead of 21 in the Rosenfeld's vectorial version\cite{sears03}. In this respect also, the Kierlik-Rosinberg version of FMT appears much simpler to manipulate and  will be more efficient in  Newton-like minimization schemes.


\end{document}